\documentstyle[aps,multicol,prl,times,tabularx,rotate,epsf]{revtex}
\begin{document}
%\draft
%\preprint{}
\title{Relaxation of a Moving Contact Line and Landau--Levich Effect}
\author{Ramin Golestanian$^{1,2,3}$ and Elie Rapha\"el$^{1}$}
\address{$^{1}$ Laboratoire de Physique de la Matiere Condensee,
College de France, URA No. 792 du CNRS, 11 place Marcelin-Berthelot, 
75231 Paris Cedex 05, France\\ 
$^{2}$Institute for Theoretical Physics, University of California, 
Santa Barbara, CA 93106-4030, USA \\
$^{3}$Institute for Advanced Studies in Basic Sciences, Zanjan 45195-159, Iran}
\date{\today}
\maketitle
\begin{abstract}
The dynamics of the deformations of a moving contact line is formulated. 
It is shown that an advancing contact line relaxes more quickly as compared
to the equilibium case, while for a receding contact line there is a corresponding 
slowing down. For a receding contact line on a heterogeneous solid surface, it is 
found that a roughening transition takes place which formally corresponds to the onset
of leaving a Landau--Levich film.
\end{abstract}
\begin{multicols}{2}%\narrowtext
When a drop of liquid spreads on a solid suface, the {\it contact line}, which is the 
common borderline between the solid, the liquid, and the corresponding
equilibrium vapor, undergoes a rather complex dynamical behavior. This dynamics
is determined by a subtle competition between the mutual interfacial energetics of the 
three phases, dissipation and hydrodynamic flows in the liquid, and the geometrical or 
chemical irregularities of the solid surface \cite{dG1}.

A most notable feature of contact lines, which is responsible for their novel
dynamics, is their anomalous elasticity as noticed by Joanny and de Gennes \cite{JdG1}. 
For length scales below the capillary 
length, which is usually of the order of 1 mm, a contact line deformation of wavevector 
$k$, denoted as $h(k)$ in Fourier space, will distort the surface of the liquid over a 
distance $|k|^{-1}$. Assuming that the surface deforms instantaneously in response to the 
contact line, the elastic energy cost for the deformation can be calculated from the 
surface tension energy stored in the distorted area, and is thus proportional to $|k|$, 
namely $E_{\rm cl}={\gamma \theta^2 \over 2} \int {d k \over 2 \pi} |k| |h(k)|^2$, in 
which $\gamma$ is the surface tension and $\theta$ is the contact angle \cite{JdG1}.

The anomalous elasticity leads to interesting {\it equilibrium} dynamics, corresponding to
when the contact line is perturbed from its static position, as studied by de Gennes 
\cite{dG2}. Balancing ${d E_{\rm cl} \over d t}$ and the dissipation, which for small contact 
angles is dominated by the hydrodynamic dissipation in the liquid nearby the contact line, 
he finds that each deformation mode relaxes to equilibrium with a characteristic frequency 
(inverse decay time) $\omega(k)=c |k|$, in which $c={\gamma \theta^3/(3 \eta \ell)}$ 
where $\eta$ is the viscosity of the liquid and $\ell$ is a logarithmic 
factor of order unity \cite{dG2}. The relaxation is thus characterized by a dynamic exponent $z$,
defined via $\omega(k) \sim |k|^z$, which is equal to 1. The linear dispersion relation implies 
that a deformation in the contact line will {\it decay} and {\it propagate} at a constant velocity 
$c$, as opposed to systems with normal line tension elasticity, where the decay and the propagation
are governed by diffusion. This behavior has been observed, and the linear dispersion relation 
has been precisely tested, in a recent experiment by Ondarcuhu and Veyssie \cite{exp1}.

As an interesting example for {\it nonequilibrium} cases, corresponding to when there is
an overall relative motion between the liquid and the solid, Landau and Levich studied 
the wetting of a plate vertically withdrawn from a completely wetting liquid at a constant 
velocity $-v$ \cite{LL}. The case of partial wetting, where the liquid has a finite contact 
angle $\theta_e$ at equilibrium, has been studied by de Gennes \cite{dG3,Voinov}. 
He finds that the velocity $v$ and the dynamic contact angle $\theta$ are related as
$v=c (\theta_e^2/\theta^2-1)/2$, and thus argues that a steady state is achieved in which
the liquid will partially wet the plate with a nonvanishing dynamic contact angle 
for pull-out velocities less than $c$, while a macroscopic Landau--Levich liquid film, 
formally corresponding to a vanishing $\theta$, will remain on the plate for higher velocities, 
as depicted in Fig.~1 \cite{dG3}. Note that at the transition there is a jump in  
the ``order parameter'' $\theta$, from $\theta_e/\sqrt{3}$ to zero.

In the presence of defects and heterogeneities in the substrate, 
which could be due to (surface) roughness or chemical 
contamination, a contact line may become {\it rough} because it 
locally deforms so as to find the path with optimal {\it pinning} 
energy \cite{dG1}. This is in contrast to the case of a perfect 
solid surface, where the contact line is {\it flat}. The roughness 
can be characterized as a scaling law that relates the statistical 
width $W$ of the contact line to its length $L$, via $W \sim 
L^{\zeta}$, with the so-called roughness exponent $\zeta$ being 
equal to $1/3$ for the case of surface disorder with short-range 
correlations \cite{Huse}. The contact line is also pinned by the 
defects, which means that a nonzero (critical) force is necessary 
to set the contact line to motion, through a depinning transition 
\cite{depinning}. It is also important to note that there may 
be numerous metastable states for the contact line due to the 
random disorder, leading to hysteresis in the contact angle 
\cite{JdG1,RJ1}. 
  
Here we study the dynamics of the deformations of a nonequilibrium (moving) contact line. 
Using a balance between the hydrodynamic dissipation in the deformed moving liquid
wedge and the rate of change of the interfacial energies, we find that
the dynamical relaxation is described by
\begin{eqnarray} 
\partial_t h(k,t)&=&-(c-v) |k| h(k,t)  \label{relax} \\
&&-{1 \over 2} \int {d q \over 2 \pi} \lambda(q,k-q) h(q,t) h(k-q,t),\nonumber 
\end{eqnarray}
in which $v$ is the average velocity of the contact line, and $\lambda(q,k-q)=
-(2 v-c) q (k-q)+{3 c} |q||k-q|+{(c-v)}|k|(|q|+|k-q|-|k|)$ 
is the leading nonlinear correction. As compared to the equilibrium case ($v=0$), 
we thus find that 
the linear relaxation is faster for an {\it advancing} contact line ($v < 0$), while it is 
slower for a {\it receding} one ($v > 0$). In particular, for a contact line that is receding at 
the ``terminal'' velocity $v=c$, which coincides with the onset of leaving a Landau--Levich film, 
linear relaxation becomes infinitely slow and the dominant relaxation is thus governed by 
the nonlinear terms. 

We also take into account the effect of surface disorder on the moving contact line dynamics,
which appears as a stochastic term in the right hand side of Eq.(\ref{relax}),
and attempt to systematically study the dynamical phase transition using renormalization
group (RG) techniques. We find that the onset of leaving a Landau--Levich film
formally corresponds to a {\it roughening} transition of the (receding) contact line,
which for a random substrate with strength $g$ takes place at a critical velocity below $c$, 
corresponding to a dynamic contact angle
\begin{equation}
{\theta_c \over \theta_e}={1 \over \sqrt{3}}
+{(11 \pi/3)^{1/3}\over 2 \sqrt{3}} \left(g \over \gamma \theta_e^2\right)^{2/3},\label{thetac}
\end{equation}       
to the leading order. We combine our results with studies of the contact line depinning 
transition \cite{depinning,RJ1}, and propose a phase diagram for the system as depicted 
in Fig.~2. In particular, we find that the phase boundaries corresponding to the dynamical
phase transition and to the depinning transition meet at a {\it triple} point, and
suggest that for stronger disorder a receding contact line will leave a Landau--Levich 
film immediately after depinning.

Let us assume that the contact line is oriented along the $x$-axis, and is moving in 
the $y$-direction with the position described by $y(x,t)=v t+h(x,t)$, as depicted in 
Fig.~3. If a line element of length $d l=d x \sqrt{1+(\partial_x h)^2}$ is displaced 
by $\delta y(x,t)$, the interfacial energy will be locally modified by two contributions: 
(i) the swept area in which liquid is replaced by vapor times the difference between 
the solid-vapor $\gamma_{SV}$ and the solid-liquid $\gamma_{SL}$ interfacial energies, 
namely, $(\gamma_{SV}-\gamma_{SL}) d l \delta y/\sqrt{1+(\partial_x h)^2}$, and 
(ii) the work done by the surface tension force, whose direction is along
the unit vector ${\hat {\bf T}}$ that is parallel to the liquid-vapor interface at 
the contact and perpendicular to the contact line, as 
$\gamma {\hat {\bf T}} \cdot {\hat {\bf y}} d l \delta y$. Note 
that we are interested in length scales below the capillary length, where gravity
does not play a role. 
The overall change in the interfacial energy of the system can thus be written as 
\begin{equation}
\delta E=\int d x \sqrt{1+(\partial_x h)^2} \left[\frac{\gamma \cos \theta_e}
{\sqrt{1+(\partial_x h)^2}}-\gamma \cos \alpha \right]\delta y(x,t),\label{deltaE}
\end{equation}                                        
in which $\alpha(x,t) \equiv \cos^{-1}\left({\hat {\bf T}} \cdot {\hat {\bf y}}\right)$,
and we have made use of the Young equation: $\gamma_{SV}-\gamma_{SL}=\gamma \cos \theta_e$.
Note that both ``forces'' should be projected onto the $y$-axis when calculating
the work done for a displacement in this direction. For small contact angles and deformations
one obtains $\sqrt{1+(\partial_x h)^2} \cos \alpha \simeq 1-\theta(x,t)^2/2$, where $\theta(x,t)$
is the local contact angle.

To calculate the dissipation, we assume that the contact angle is sufficiently small 
so that the dominant contribution comes from the viscous losses in the hydrodynamic flows 
of the liquid wedge \cite{dG1}. For a slightly deformed contact line, we assume that the 
dissipation can be approximated by the sum of contributions from wedge-shaped slices with 
local contact angles $\theta(x,t)$. This is a reasonable approximation because most of the 
dissipation is taking place in the singular flows near the tip of the wedge \cite{dG1,dG2}. 
Using the result for the dissipation in a perfect wedge which is based on the lubrication 
approximation \cite{dG1,Huh}, we can calculate the total dissipation as \cite{dG2}
\begin{equation}
P=\int d x \sqrt{1+(\partial_x h)^2}\;
\left\{3 \eta \ell \left[v+\partial_t h(x,t)\right]^2\over \theta(x,t)\right\} .\label{P1}
\end{equation}      
We can now use Eqs.(\ref{deltaE}) and (\ref{P1}) to calculate $-{d E \over d t}$ and $P$, 
and set them equal to each other to derive the dynamical equation. It yields 
\begin{equation}
3 \eta \ell \left[v+\partial_t h(x,t)\right] ={\gamma \over 2} 
{\theta(x,t)\left(\theta_e^2-\theta(x,t)^2\right) \over \sqrt{1+(\partial_x h)^2}}.\label{vtheta}
\end{equation}
The above equation might simply be recovered by locally applying the 
result of Ref.\cite{dG2} for straight contact lines, with the additional geometrical 
factor, which is needed when the direction of motion is not perpendicular to the 
contact line, taken into account.  

To complete the calculation, we need to solve for the profile of 
the surface and the corresponding angles as a function of $h$. 
One can show that the surface profile $z(x,y)$ near the contact 
line can be found as a solution of the Laplace equation 
$(\partial^2_x+\partial^2_y)z(x,y)=0$, so as to minimize the surface area.
The solution reads 
$z(x,y)=\theta y+\int {d k \over 2 \pi} \beta(k) \exp\left(i k x-|k| y\right)$,
where $\beta(k)=-\theta \left[h(k)+\int {d q \over 2 \pi} |q| h(q) h(k-q)+O(h^3)\right]$
is found using the boundary condition $z(x,h(x))=0$ \cite{JdG1}. 
We therefore find $\theta(x)=\theta [1+\int {d k \over 2 \pi} |k| h(k) e^{i k x}
+{1 \over 2} \int {d k \over 2 \pi}{d k' \over 2 \pi} f(k,k') h(k) h(k') e^{i (k+k') x}]$
with $f(k,k')=|k+k'|(|k|+|k'|)-(k+k')^2+k k'$, which can then be used in Eq.(\ref{vtheta}) 
to yield Eq.(\ref{relax}), and the relation between $v$ and $\theta$ as described above and 
depicted in Fig.~1.

The linear relaxation of a moving contact line thus takes place at characteristic frequencies 
which obey the modified dispersion relation $\omega(k)=(c-v) |k|$, and as the onset of leaving
a Landau--Levich film corresponding to $v=c$ is approached, the relaxaxtion becomes progressively
slower.

In addition to dissipation and elasticity, the dynamics of a contact line is
also affected by the defects and heterogeneities in the substrate, which are present
in most practical cases. If the intefacial energies $\gamma_{SV}$ and $\gamma_{SL}$ are
space dependent, a displacement of the contact line is going to lead to a change in energy as 
$\delta E_{\rm d}=\int d x g(x,v t+h(x,t)) \delta y(x,t)$, where 
$g(x,y)=\gamma_{SV}(x,y)-\gamma_{SL}(x,y)-({\bar \gamma_{SV}}-{\bar \gamma_{SL}})$.
Incorporating this contribution in the force balance leads to a noise term on the right hand side 
of Eq.(\ref{relax}) of the form $\eta(x,t)={\theta \over 3 \eta \ell} g(x,vt)$ to the leading order.
Note that this is a good approximation provided we are well away from the depinning 
transition, and the contact line is moving fast enough \cite{dG1,JdG1,depinning,RJ1}.
Assuming that the surface disorder has short range correlations (so that the correlation length
is a microscopic length $a$) with a Gaussian distribution described by $\langle g(x,y) \rangle=0$ and 
$\langle g(x,y) g(x',y')\rangle=g^2 a^2 \delta(x-x') \delta(y-y')$, we can deduce the distribution
of the noise as: $\langle \eta(x,t) \rangle=0$ and $\langle \eta(x,t) \eta(x',t')\rangle
={c^2 g^2 a^2 \over \gamma^2 \theta^4 |v|}\delta(x-x') \delta(t-t')$.  

In the presence of the noise, the contact line undergoes dynamical fluctuations. These fluctuations 
can best be characterized by the width of the contact line, which is defined as 
%$W^2(L,t)\equiv {1 \over L} \int d x \langle h(x,t)^2 \rangle=
%{c^2 g^2 a^2 \over 2 \pi (c-v)\gamma^2 \theta^4 |v|} \int_{\pi/L}^{\pi/a} {d k \over k}  
$W^2(L,t)\equiv {1 \over L} \int d x \langle h(x,t)^2 \rangle$. 
Using the scaling form of the two-point correlation function, one 
can show that $W \sim t^{\zeta/z}$ for intermediate times, while it 
saturates to $W \sim L^{\zeta}$ at long times. Similarly, we can study 
the fluctuations in the order parameter field $\delta \theta(x,t)=\theta(x,t)-\theta$,
and find 
\begin{equation}
\langle \delta \theta(x,t)^2 \rangle \sim 1-B/t^{2(1-\zeta)/z},  
\label{theta0}
\end{equation}
where $B$ is a constant. Note that the order parameter fluctuations approach a finite limit at long times provided
$\zeta < 1$. 

Keeping only the linear term in Eq.(\ref{relax}), we calculate the width of the contact line as 
\begin{equation}
W(L,t) \sim \left\{\begin{array}{ll}
\sqrt{\ln\left[(c-v) t/a\right]}, &\; {a \over (c-v)} \ll t \ll {L \over (c-v)},  \\  
\sqrt{\ln\left(L/a\right)}, &\; t \gg {L \over (c-v)},
\end{array} \right. \label{W1}
\end{equation}    
and, similarly, the order parameter fluctuations as
\begin{equation}
\langle \delta \theta(x,t)^2 \rangle \sim 1-{B/t^2}, \label{theta1}
\end{equation} 
for $t \gg {a \over (c-v)}$. We thus find $\zeta=0$ and $z=1$ within the linear theory.

The nonlinear terms in Eq.(\ref{relax}) will modify the above results only when it becomes
appreciable at long length scales, as compared to the linear term. The ratio of the two terms 
in Eq.(\ref{relax}) scales as ${c (L/a)^{2 \zeta-1} \over (c-v) (L/a)^{\zeta}} \sim {a c \over L(c-v)}$, and 
is thus appreciable {\it only} when the smallest time scale in the linear theory $a/(c-v)$
becomes comparable to $L/c$. This happens near the onset of leaving a Landau--Levich film.

Let us now attempt to systematically study the dynamical phase transition, corresponding
to leaving a Landau--Levich film, using the RG scheme. 
The dynamical equation, which can be generally written in $d$ dimensions as 
$\partial_t h(k,t)=-\nu |k| h(k,t)-{1 \over 2} \int {d^d q \over 2 \pi} \lambda(q,k-q) 
h(q,t) h(k-q,t)+\eta(k,t)$, with 
$\lambda(q,k-q)=-\lambda_1 q (k-q)+\lambda_2 |q||k-q|+\lambda_3 |k|(|q|+|k-q|-|k|)$,
belongs to the general class of Kardar-Parisi-Zhang equations \cite{KPZ}. 
We take a noise spectrum given by $\langle \eta(k,t) \rangle=0$ and 
$\langle \eta(k,t) \eta(k',t')\rangle=2 D (2 \pi)^d \delta^d(k+k') \delta(t-t')$,
and employ standard RG techniques following Ref. \cite{KPZ} to calculate the RG equations 
describing the flow of the coupling constants. We find 
\begin{eqnarray}
&&{d \nu \over d l}= \nu \left[z-1-{S_d (\pi/a)^{d+1} (\lambda_1+\lambda_2)(\lambda_2+\lambda_3) 
D \over 2 (2 \pi)^d \nu^3}\right], \nonumber  \\       
&&{d \lambda(q,k-q) \over d l}= \lambda(q,k-q) \left(\zeta+z-2\right), \label{RG} \\      
&&{d D \over d l}= D \left[z-2 \zeta-d+{S_d (\pi/a)^{d+1} (\lambda_1+\lambda_2)^2 
D^2 \over 4 (2 \pi)^d \nu^3}\right], \nonumber
\end{eqnarray}
to the one-loop order, in which $S_d$ is the surface area of a unit sphere in $d$ dimensions.

To study the fixed point structure of the above set of flow equations, it is convenient to introduce 
the dimensionless coupling constant $U={S_d (\pi/a)^{d+1} (\lambda_1+\lambda_2)
(\lambda_2+\lambda_3) D/[2 (2 \pi)^d \nu^3]}$, and thus have $z=1+U$ and $\zeta=1-U$ at
the fixed points. The flow equation for $U$ reads: 
$d U/d l=-(d+1) U+\left[6+(\lambda_1+\lambda_2)/(\lambda_2+\lambda_3)\right] U^2/2$,
which has two stable fixed points at $U=0$ (linear theory) and $U=\infty$ 
(strong coupling), as well as an intermediate unstable fixed point at 
$U=U^*\equiv{2(d+1) \over 6+(\lambda_1+\lambda_2)/(\lambda_2+\lambda_3)}$. 

For $U<U^*$, the nonlinearity is irrelevant and the exponents are given by the linear theory, i.e. 
$\zeta=0$ and $z=1$, while for $U>U^*$ the behavior of the system is governed by a strong coupling fixed 
point which cannot be studied perturbatively. The fixed point at $U^*$ corresponds to a roughening transition
of the moving contact line. The exponents at the transition are  
$z=1+{2(d+1) \over 6+(\lambda_1+\lambda_2)/(\lambda_2+\lambda_3)}$ and 
$\zeta=1-{2(d+1) \over 6+(\lambda_1+\lambda_2)/(\lambda_2+\lambda_3)}$, which are
nonuniversal. The strong coupling fixed point should presumably describe 
the Landau--Levich film.

We can also study how the transition is approached by linearizing the flow equation
near the fixed point. Setting $U=U^*+\delta U$, we find $d \delta U/d l=(d+1) \delta U$
that would imply divergence of the correlation length near the transition as
$\xi \sim |\delta U|^{-\nu}$ with $\nu=1/(d+1)$. The correlation 
length corresponds to the typical size of rough segments in the 
contact line, which should diverge as the transition is approached.

Using the equation for the phase boundary $U=U^*$ with $d=1$, and the values of the 
coupling constants that we can read off from Eq.(\ref{relax}), we map out the phase diagram 
of the system as depicted in Fig.~2. In particular, we find the asymptotic form of the phase
boundary for weak disorder as reported in Eq.(\ref{thetac}) above, and the nonuniversal 
exponents $z={17 \over 11}+{15 \over 121}({11 \pi \over 3})^{1/3}({g \over \gamma \theta_e^2})^{2/3}$ 
and $\zeta={5 \over 11}-{15 \over 121}({11 \pi \over 3})^{1/3}({g \over \gamma \theta_e^2})^{2/3}$
to the leading order, and $\nu=1/2$. The order parameter fluctuations at the transition
are given by Eq.(\ref{theta0}) with the above choice for $z$ and $\zeta$, and remain finite 
since $\zeta < 1$. We also find that the triple point is located at $\theta_t/\theta_e=0.887$ 
and $g_t/(\gamma \theta_e^2)=0.138$, which is interestingly still within the weak disorder limit. 

It is important to note that since our approach is only valid when the system is away 
from the depinning transition, it is not clear that we can trust the 
prediction of our theory in the vicinity of the triple point. In 
particular, how the two phase boundaries merge at the triple 
point, and the nature of the third phase boundary (dashed line) 
remains to be clarified. Furthermore, the above calculation is 
only restricted to the case of small contact angles where the 
dissipation is dominated by the singular hydrodynamic flow near 
the tip of the liquid wedge. Should other mechanisms of dissipation
become important, the above picture would be altered \cite{Blake}. 
 
We finally mention that there could be two types of experiments to check these results,
which we hope to motivate. The first type would be an analogue of the
Ondarcuhu--Veyssie experiment \cite{exp1}, which could probe the relaxation of a moving 
contact line and measure the velocity dependence of the dispersion relation. 

The second type of experiments would correspond to a systematic 
study of the onset of leaving a Landau--Levich film for receding 
contact lines on a disordered substrate. In particular, it would be interesting 
to look for a roughening of the contact line before the Landau--Levich film is formed.

%\acknowledgments

We are grateful to J. Bico, R. Bruinsma, P.G. de Gennes, M. Kardar, and D. Qu\'er\'e for invaluable 
discussions and comments. This research was supported in part by the National Science Foundation under 
Grants No. PHY94-07194 and DMR-98-05833.

\begin{figure}
%\centerline{\epsfxsize 5.6cm \rotatebox{-90}{\epsffile{}}}
%\centerline{\epsfxsize 4cm \rotate[r]{\epsffile{ev.eps}}} 
\centerline{\epsfxsize 6.0cm {\epsffile{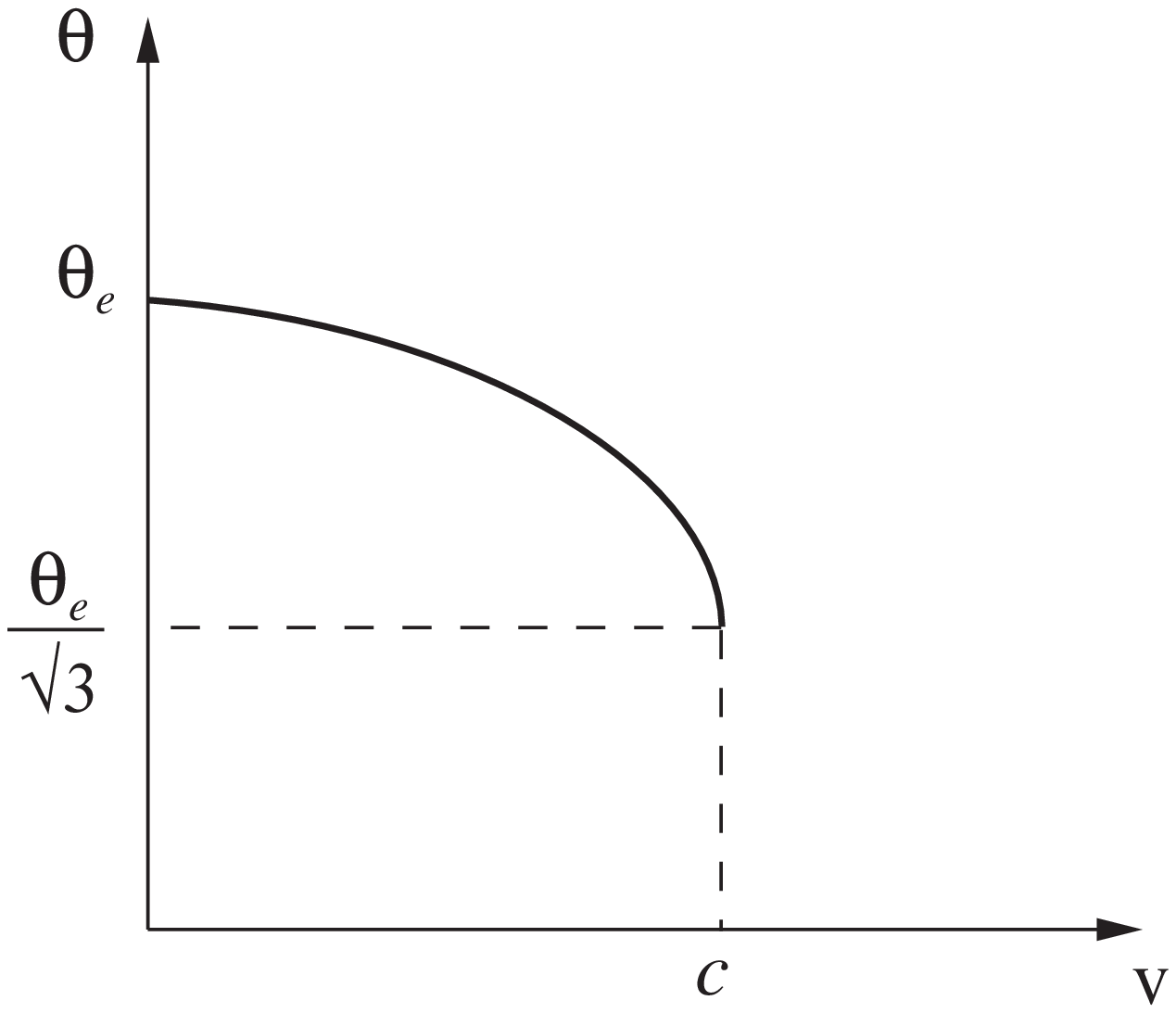}}}   
\vskip .1truecm
%\caption{}
FIG.~1. Dynamic contact angle as a function of pull-out velocity \cite{dG3}. 
\label{fig1}
\end{figure}         

\begin{figure}
%\centerline{\epsfxsize 5.6cm \rotatebox{-90}{\epsffile{}}}
\centerline{\epsfxsize 7.0cm {\epsffile{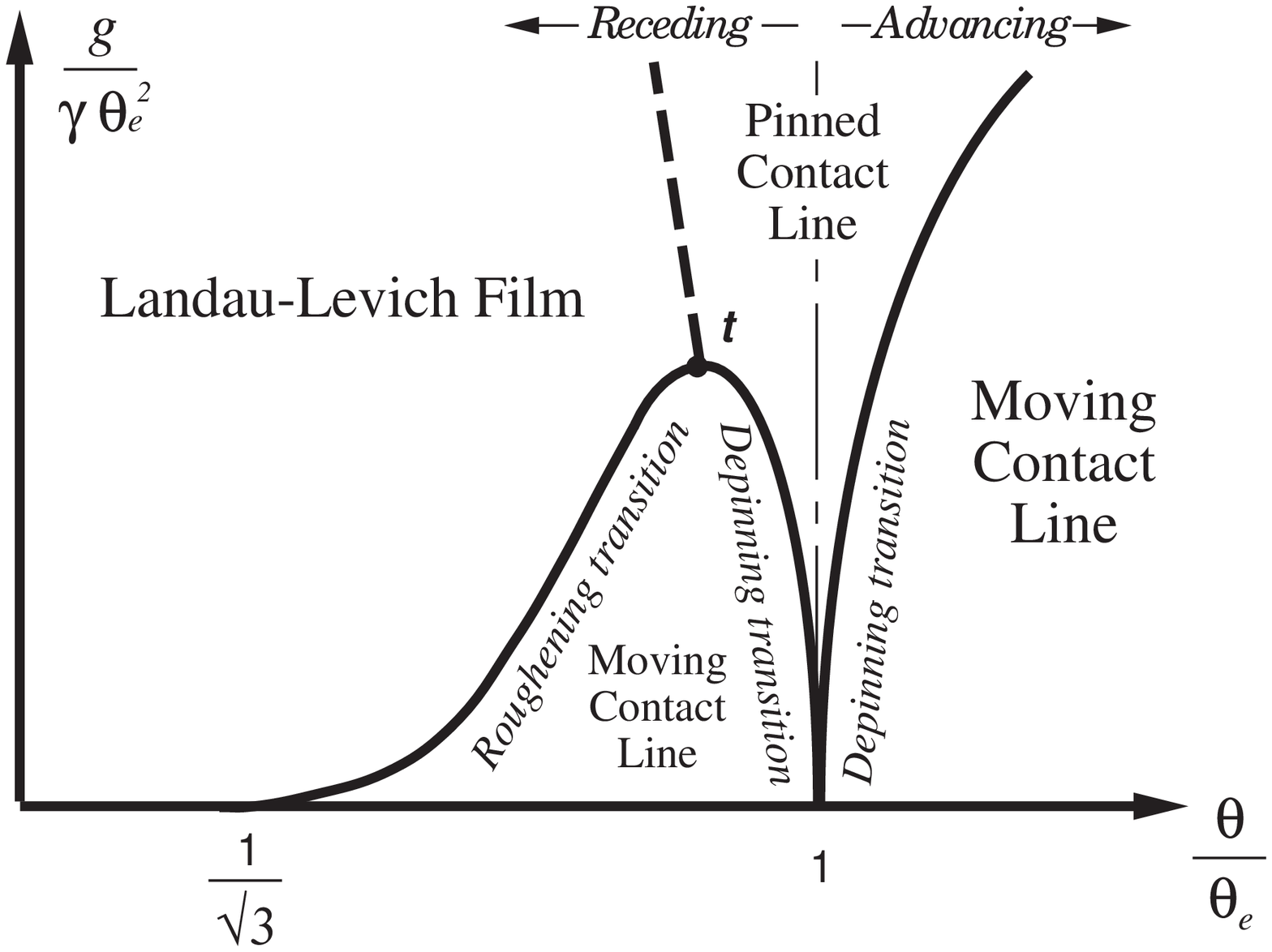}}}
\vskip .1truecm
%\caption{}
FIG.~2. The suggested phase diagram of a contact line on a disordered substrate. The depinning
transition line is taken from Ref.\cite{RJ1}, corresponding to the receding and the advancing
contact angles. The asymptotic form for the roughening transition line is given in 
Eq.(\ref{thetac}).  
\label{fig2}
\end{figure}          
            
\begin{figure}
%\centerline{\epsfxsize 5.6cm \rotatebox{-90}{\epsffile{}}}
\centerline{\epsfxsize 7.0cm {\epsffile{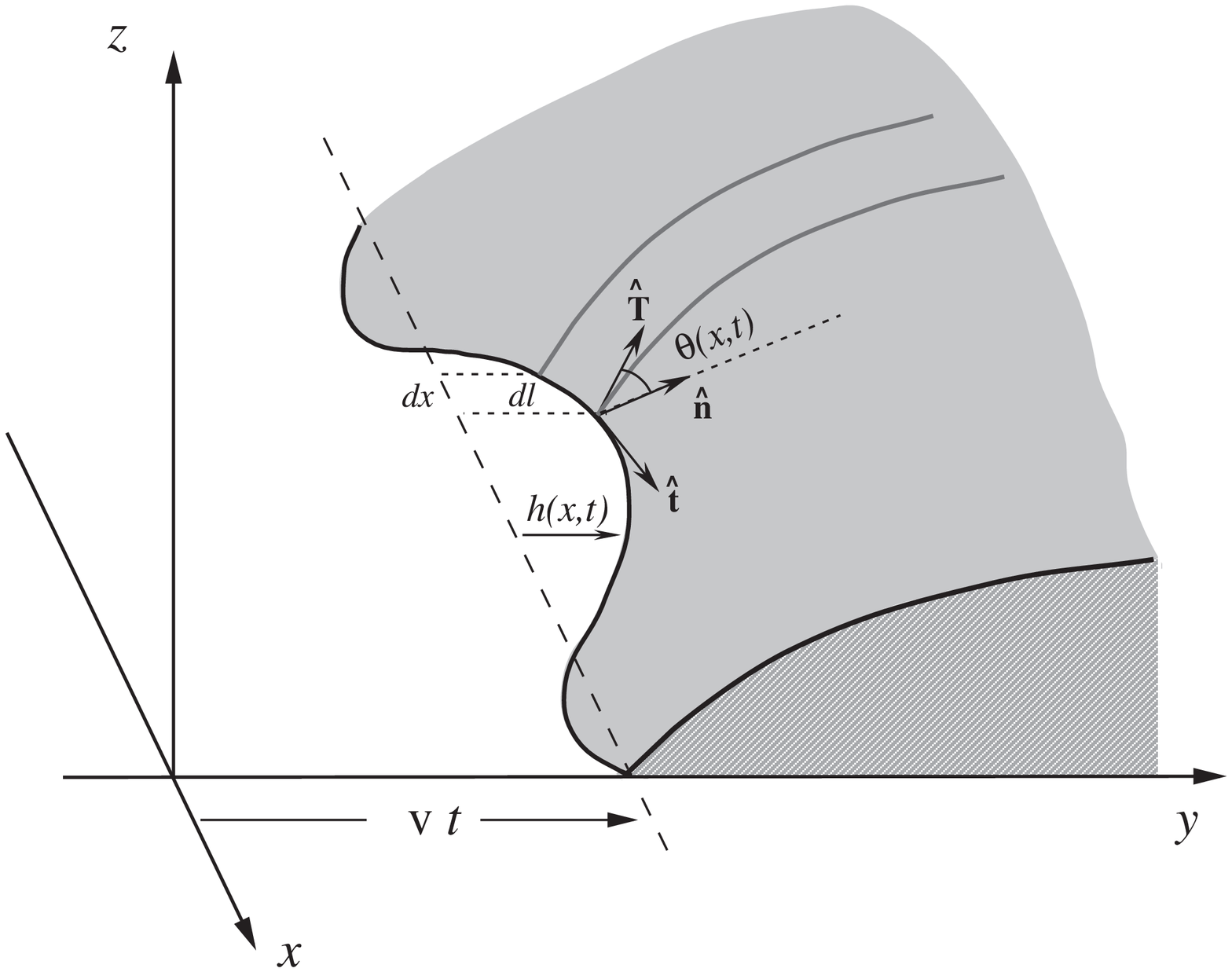}}}
\vskip .1truecm
%\caption{}
FIG.~3. The schematics of the system. 
\label{fig3}
\end{figure}      
  
\end{multicols}
\end{document}